\documentclass{PoS}

\def\be{\begin{equation}}
\def\ee{\end{equation}}
\def\bea{\begin{eqnarray}}
\def\eea{\end{eqnarray}}
\def\ben{\begin{enumerate}}
\def\een{\end{enumerate}}
\usepackage{wrapfig}

\title{Insisting on the role of experimental data: the pseudoscalar-pole piece to the $(g_\mu-2)$ and the $|V_{ub}|$ from $B \to \pi \ell \nu_{\ell}$ differential branching ratio}

\ShortTitle{Insisting on the role of experimental data}

\author{Sergi Gonz\`{a}lez-Sol\'{i}s\\
CAS Key Laboratory of Theoretical Physics, Institute of Theoretical Physics, Chinese Academy of Sciences, Beijing 100190, China\\
\email{sgonzalez@itp.ac.cn}}

\author{\speaker{Pere Masjuan}\thanks{The work of S.G-S is supported in part by the CAS President's International Fellowship Initiative for Young International Scientists (Grant No. 2017PM0031), by the Sino-German Collaborative Research Center \textquotedblleft Symmetries and the Emergence of Structure in QCD\textquotedblright\,(NSFC Grant No. 11621131001, DFG Grant No. TRR110), and by NSFC (Grant No. 11647601); the work of P.M by CICYTFEDER-FPA2014-55613-P, 2014-SGR-1450, the CERCA Program/Generalitat de Catalunya,  the Secretaria d'Universitats i Recerca del Departament d'Empresa i Coneixement de la Generalitat de Catalunya; and the work of P.S-P by the Czech Science Foundation (grant no. GACR 15-18080S).}\\
        Grup de F\'{\i}sica Te\`orica, Departament de F\'{\i}sica, 
  Universitat Aut\`onoma de Barcelona,  and Institut de F\'{\i}sica d'Altes Energies (IFAE), The Barcelona Institute of Science and Technology (BIST), 
  Campus UAB, E-08193 Bellaterra (Barcelona), Spain\\
        E-mail: \email{masjuan@ifae.es}}

\author{Pablo Sanchez-Puertas\\
Faculty of Mathematics and Physics, Institute of Particle and Nuclear Physics, Charles University in Prague, V Hole\v{s}ovi\v{c}k\'ach 2, Praha 8, Czech Republic\\
\email{sanchezp@ipnp.troja.mff.cuni.cz}
}

\abstract{We employ a mathematical framework based on rational approximants in order to calculate meson form factors. The method profits from unitary, is systematic and data based, and is able to ascribe a systematic uncertainty which provides for the desired model independence. Two examples are discussed: the transition form factor entering the pseudoscalar-pole piece of the hadronic light-by-light contribution to the anomalous magnetic moment of the muon, and the $B \to \pi$ form factor participating the $B\to\pi\ell\nu_{\ell}$ differential branching ratios which allows to determine the $|V_{ub}|$ CKM parameter.}

\FullConference{XVII International Conference on Hadron Spectroscopy and Structure - Hadron2017\\
		25-29 September, 2017\\
		University of Salamanca, Salamanca, Spain}

\begin{document}

\section{Introduction}

Hadronic form factors (FF) play an important role in multiple and different scenarios in particle physics. An accurate description in both space-like (SL) and time-like (TL) regions is an important task with large benefits starting from the direct study of experimental data up to complicated calculations at the frontier of the Standard Model. A complete model-independent description of both SL and TL would demand a full knowledge of QCD in both its perturbative and non-perturbative regimes, knowledge not yet aquired. An alternative to such enterprise would pursuit a synergy between theory and experiment, between the formal calculation and the experimental data. In this respect, one would direct oneself towards a model-independent and data-driven phenomenological description of FF. 

This synergy should develop a method as simple as possible, maximally transparent, fully satisfying the analyticity and unitarity of the FF.  If possible, the method should not use any assumption, only approaches, improvable without \textit{ad hoc} statements. It shall provide a systematic method as well, and in two different senses: easy to update whenever new experimental data or new theoretical calculations are available; capable of providing a purely theoretical error from the approaches performed. Finally, should be predictive and checkable.

This catalogue of wishes can be addressed within the Theory of Pad\'e approximants (PA). The connexion with the mathematical problem is given by the well defined \textit{general rational Hermite interpolation problem}. This problem corresponds with the situation where a function should be approximated but previous information about it is scarce and spread over certain information on a given set of points together with a set of derivatives. 

Analyticity and unitary of vector FFs imply them to be Stieltjes\footnote{A function $f(z)$ is called Stieltjes if obeys a dispersion relation given in terms of a positive definite spectral function.} functions~\cite{Pade,Masjuan:2009wy,Works}. As such, any diagonal or subdiagonal PA sequence, on top of converging, must have all its poles lying on the real axis, an extremely useful feature when analyzing experimental data. In this case, PAs act as the \textit{guarantor} of unitary through its pole position after the fit to experimental data. In case complex-conjugate poles or \textit{defects} (a pole with a close-by zero almost canceling each other) appear~\cite{Masjuan:2007ay}, since they are not allowed by the convergence theorems~\cite{Masjuan:2009wy}, they are a clear indication that experimental data are not satisfying FF's unitary. Thus, a bootstrap method will allow to pinpoint the cause of the violation of unitary and via neglecting the identified experimental datum, the fit is immediately improved.  This method has been successfully used already to study pseudsocalar transition form factors (TFF) and explore their role in extracting low-energy parameters, the $\eta-\eta'$ mixing angle, and parameterizations of the doubly virtual TFF, as well as for the pseudoscalar contribution to the HLBL of the muon $(g-2)$ (for a recent review, see Ref.~\cite{review,Masjuan:2017tvw}). 

In this letter, the PA method is presented to discuss the role of experimental data in: $i)$ the $\pi^{0},\eta$, $\eta^{\prime}$ SL TFFs and the TL TFF entering the description of the $\pi^{0},\eta$ and $\eta^{\prime}$ Dalitz decays; $ii)$ the TL $B \to \pi$ semileptonic FF appropriate to extract the $|V_{ub}|$ CKM parameter.

First, some boring maths to show how unitary constraints arise for PA sequences. Let us consider the power series expansion of a function $f(z)$ around the origin on the complex plane $(z\to0)$ as $f(z)=\sum_{n=0}^{\infty}c_{n}z^{n}$ with a certain radius of convergence. Strictly speaking a PA to $f(z)$ is a polynomial of order $N$ over a polynomial of order $M$
\begin{equation}
P^{N}_{M}(z)=\frac{\sum_{j=0}^{N}a_{j}(z)^{j}}{\sum_{k=0}^{M}b_{k}(z)^{k}}=
\frac{a_{0}+a_{1}z+\cdots+a_{N}(z)^{N}}{1+b_{1}z+\cdots+b_{M}(z)^{M}}\ ,
\label{PadeApproximant}
\end{equation}
constructed such that\footnote{With any loss of generality, we take $b_{0}=1$ for definiteness.} its coefficients satisfy the accuracy-through-order conditions with $f(z)$, this is the Taylor expansion of $P_{M}^{N}(z)$ matches the series $f(z)$ up to the highest possible order 
\begin{equation}\nonumber
f(z)-P_{M}^{N}(z)=\mathcal{O}(z)^{M+N+1}\,.
\label{condition}
\end{equation}

\begin{wrapfigure}{l}{0.45\textwidth}
\centering
\includegraphics[width=0.4\textwidth,trim=10.7cm 8.5cm 11cm 8.5cm, clip=true]{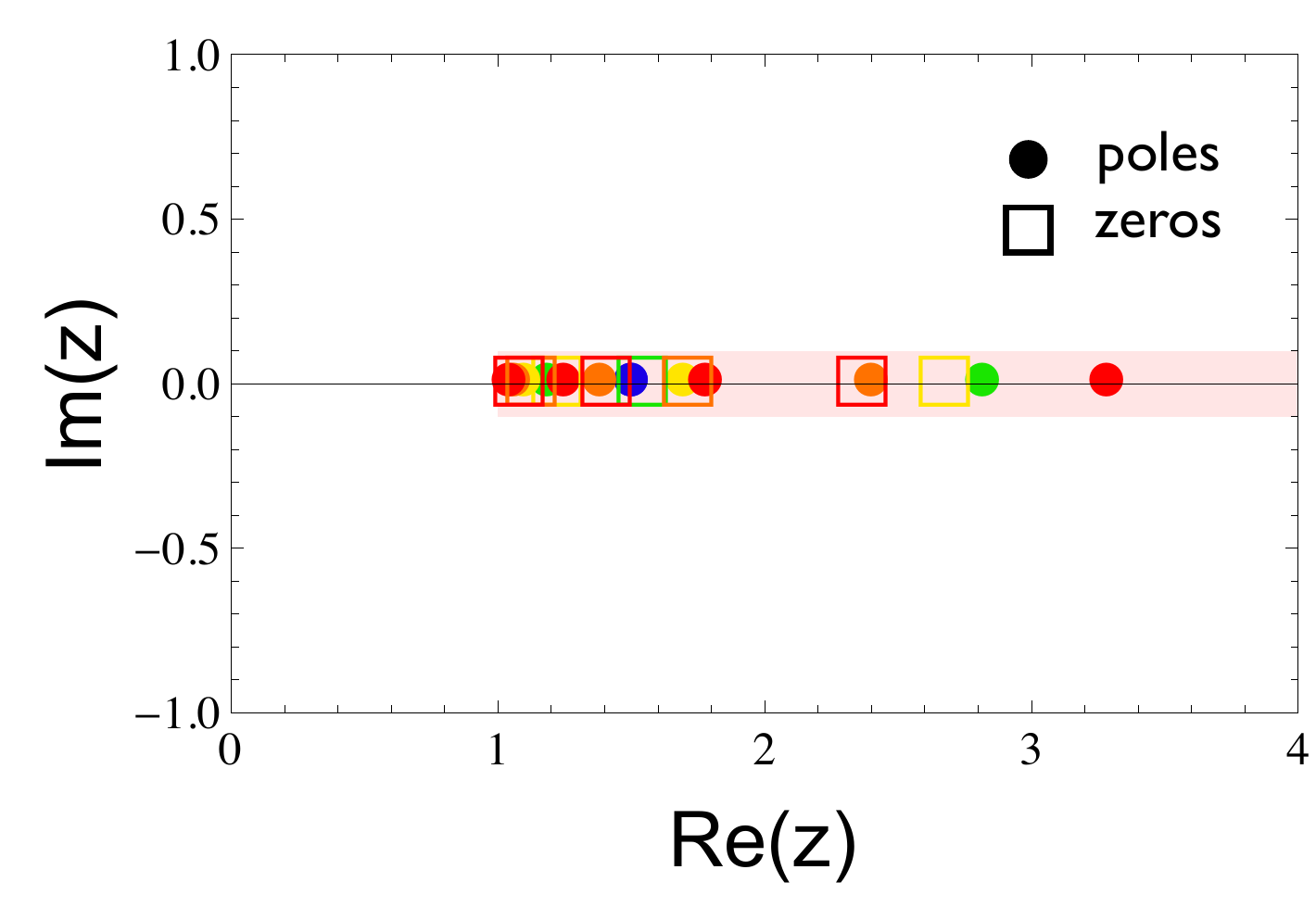}
\caption{Poles (dotes) and zeros (empty squares) of the $P^N_N$ approximants from the $\frac{1}{z}{\rm{ln}}(1+z)$ function for $N$ up to $4$.} 
\vspace{-0.8cm}
\label{PadeLog}
\end{wrapfigure}

For example, let us consider 
\begin{equation}
f(z)=\frac{1}{z}{\rm{ln}}(1+z)\, =\sum_{n=0}\frac{(-z)^{n}}{n+1}=1-\frac{z}{2}+\frac{z^{2}}{3}-\frac{z^{3}}{4}+\frac{z^{4}}{5}-\frac{z^{5}}{6}+\mathcal{O}(z^{6})\,,
\label{logfunctionexample}
\end{equation}
which converges for $|z|<1$ and diverges elsewhere. To determine the $P_{1}^{0}(z)$ we expand it in a Taylor series
\begin{equation}\nonumber
P_{1}^{0}(z)=\frac{a_{0}}{1+b_{1}z}=a_{0}-a_{0}b_{1}z+\mathcal{O}(z^{2})\,,
\end{equation}
and compare with $f(z)=1-z/2+\mathcal{O}(z)$ to get $a_0$ and $b_1$, and so on for higher $P^N_N(z)$ and $P^N_{N+1}(z)$ approximants. If they converge, their poles must lie in the real axes since $f(z)$ is a Stieltjes function. As a matter of example, in Fig.~\ref{PadeLog} we provide a graphical account of poles and zeros of 
\begin{equation}\nonumber
P_{1}^{0}(z)=\frac{1}{1+\frac{z}{2}}\, \,, P_{1}^{1}(z)=\frac{1+\frac{z}{6}}{1+\frac{2z}{3}}\, \,,P_{2}^{1}(z)=\frac{1+\frac{z}{2}}{1+z+\frac{z^{2}}{6}}\, \,, P_{2}^{2}(z)=\frac{1+\frac{7z}{10}+\frac{z^{2}}{30}}{1+\frac{6z}{5}+\frac{3z^{2}}{10}}\,,
\end{equation}
to the function Eq.~(\ref{logfunctionexample}).

\vspace{-0.5cm}
The sequence converges rapidly even beyond the convergence's radius $|z|<1$. In case the poles would lie outside the physical cut, this would imply the coefficients $c_n$ to have some noise, some error detected by the PAs and easy to remove!

\section{Pseudoscalar transition form factor}

\begin{wrapfigure}{l}{0.45\textwidth}
\includegraphics[width=0.4\textwidth]{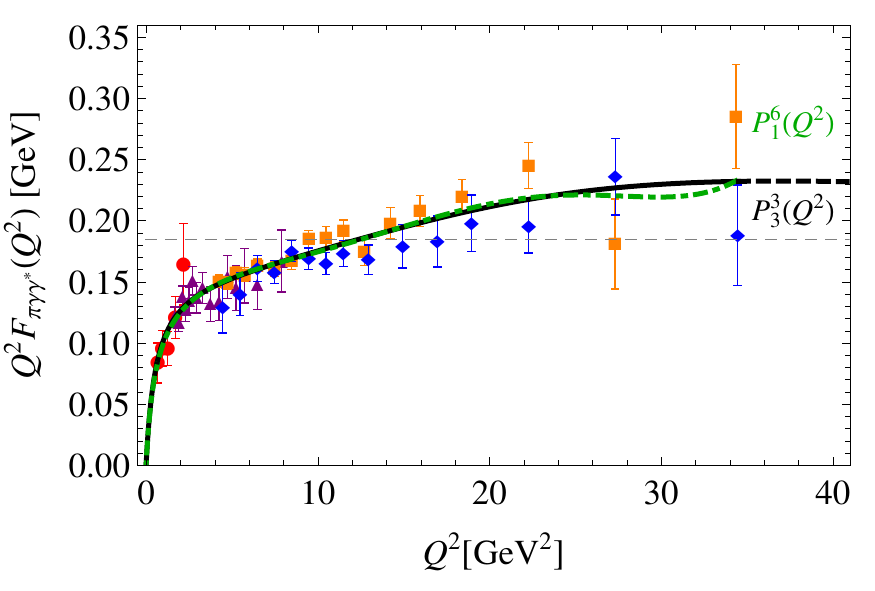}
\includegraphics[width=0.4\textwidth]{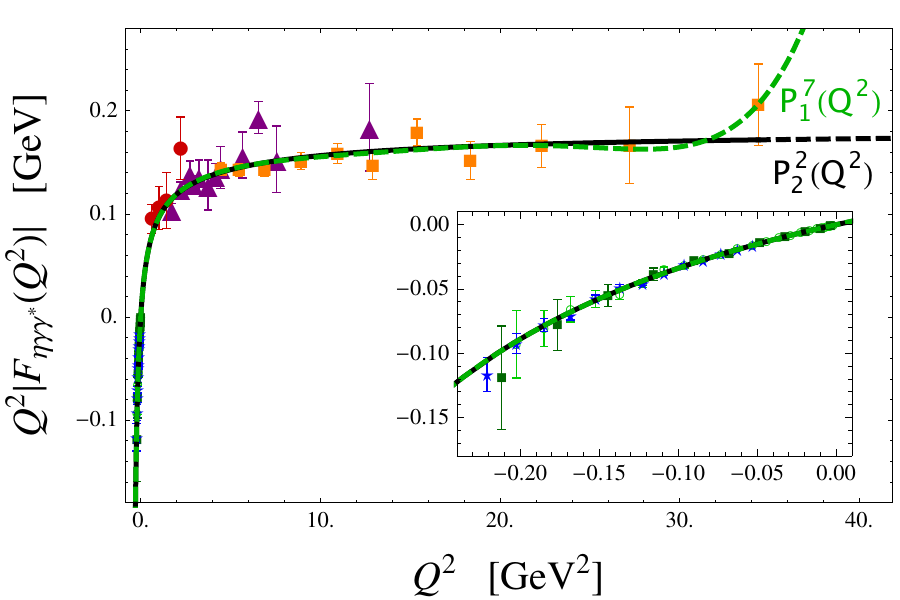}
\includegraphics[width=0.415\textwidth]{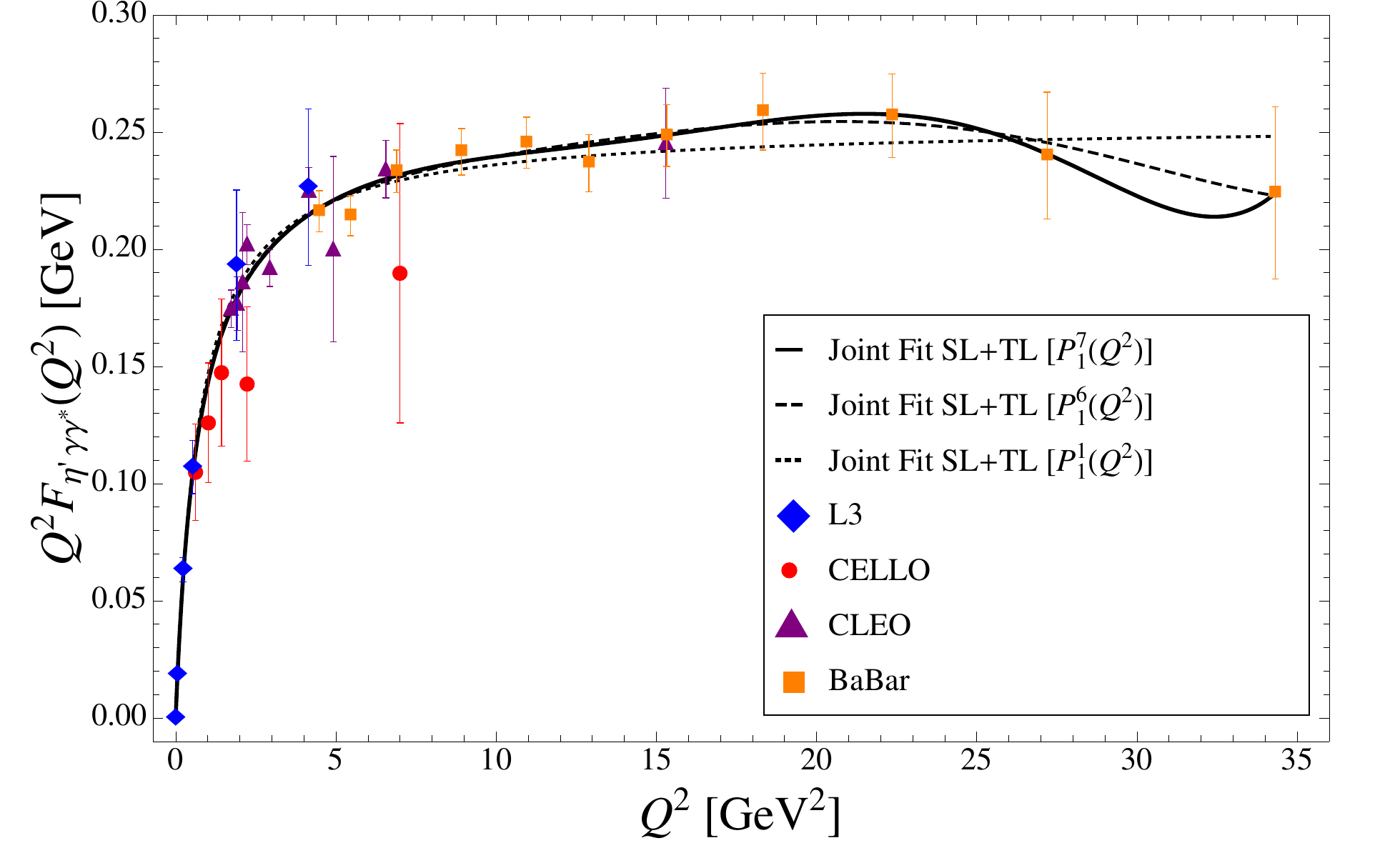}
\caption{$\pi^0$ (upper panel), $\eta$ (middle panel), and $\eta^\prime$ (lower panel) TFFs. Our best $P^L_{1}(Q^2)$ and $P^N_{N}(Q^2)$ fits are shown. Dashed lines display the extrapolation of the $P^N_{N}(Q^2)$ at $Q^2=0$ and $Q^2\to\infty$. Data from CELLO, CLEO, {\textit{BABAR}}, BELLE, and L3~\cite{sldata} for SL, and NA60, A2, BESIII~\cite{tldata} for TL. Figures form Refs.~\cite{Works}.}
\vspace{-1.5cm}
\label{fig1}       
\end{wrapfigure}

The TFF describes the effect of the strong interaction on the $\gamma^*\gamma^* - P$ transition (where $P=\pi^0, \eta, \eta^\prime\cdots$) and is represented by a function $F_{P\gamma^*\gamma^*}(q_1^2,q_2^2)$ of the photon virtualities $q_1^2$, and $q_2^2$. The singly virtual TFF, which depends on the transferred momentum of the virtual photon, has a well defined unitary cut in the TL region starting at the $2m_{\pi}^2$ threshold and no singularities in the SL region since it is a Stieltjes function~\cite{Works}. The nature of the production threshold is of vector type, which guarantees a very smooth off-set of the imaginary part of the threshold discontinuity in the TL. The experimental information on the TFFs together with the theoretical knowledge on their kinematic limits yield the opportunity for a nice synergy between experiment and theory in a simple, easy, systematic, and user-friendly way. 

We proposed in Refs.~\cite{Works} to use a sequence of PA  to fit together SL~\cite{sldata} and TL~\cite{tldata} data. Since PAs  are constructed from the Taylor expansion of the $F_{P\gamma^*\gamma}(Q^2)$, from the fits we can obtain the derivatives of the $F_{P\gamma^*\gamma}(Q^2)$ defined as~\cite{Works}:
\begin{equation}\label{Taylor}
F_{P\gamma*\gamma} (Q^2)=a_0^P\left(1+b_{P} \frac{Q^2}{m_{P}^2}+c_{P} \frac{Q^4}{m_{P}^4}+ \dots\right)\, ,
\end{equation}
\noindent
where $a_0^P$ is related to $P\to \gamma\gamma$, $b_P$ and $c_P$ are the slope and curvature resp., fundamental quantities for constraining models to evaluate hadronic contributions. Our results are collected in Table~\ref{tab1}.

In case complex-conjugated poles in our approximants would appear, that would be a clear indication of an underestimation of experimental errors as we discuss in the next section.
\begin{wraptable}{l}{0.62\textwidth}
\vspace{-0.5cm}
\centering
\caption{$\pi^0,\eta$, and $\eta^\prime$ slope $b_P$, curvature $c_P$, and asymptotic limit ($Q^2 \to \infty$) from Ref.~\cite{Works}.}
\label{tab1}       
\begin{tabular}{cccc}
\hline
 & $b_P$ & $c_P$ & $\lim_{Q^2\to\infty }Q^2F_{P\gamma^*\gamma}(Q^2) $    \\\hline
$\pi^0$ & $0.0324(22)$ & $1.06(27)\cdot 10^{-3}$ & $2 f_{\pi} $   \\
$\eta$ & $0.576(11)$ & $0.339(15)$ & $0.177(20) \textrm{GeV}$ \\ 
$\eta^\prime$ & $1.31(4)$ & $1.74(9)$ & $0.254(3) \textrm{GeV}$  \\ \hline
\end{tabular}
\end{wraptable}

FF are not interesting by themselves as represent the knowledge of QCD in a nutshell, but also for their important role on precision calculations of low-energy Standard Model observables such as the anomalous magnetic moment of the muon. With this method, we updated in~\cite{review,Masjuan:2017tvw} the HLBL to $(g_\mu-2)$ to be $a_{\mu}^{{\mathrm{HLBL}}} = (12.1\pm1.5)\times10^{-10}$. FF are important for $P \to \ell \ell$~\cite{Masjuan:2015lca} as well. Both showing deviations between theory and experiment point towards a search of New Physics.

\vspace{-0.cm}
\section{Study of $B\to\pi\ell\nu_{\ell}$ and $B^{+}\to\eta^{(\prime)}\ell^{+}\nu_{\ell}$ decays and determination of $|V_{ub}|$}

$|V_{ub}|$ is one of the least known entries of the CKM matrix. It is typically determined from inclusive and exclusive semileptonic decays through $B\to X_{u}\ell\nu_{\ell}$ and $B\to\pi\ell\nu_{\ell}$, respectively. The  PDG reported values~\cite{Agashe:2014kda} showed a $2.4\sigma$ deviation between the inclusive, $|V_{ub}|=(4.49\pm0.15^{+0.16}_{-0.17}\pm0.17)\cdot10^{-3}$, and the exclusive, $|V_{ub}|=(3.70\pm0.10\pm12)\cdot10^{-3}$ \cite{Vubpdg2014} determinations.

The origin of this discrepancy is still unclear and any combined average must be borrowed with caution~\cite{Agashe:2014kda}.
The exclusive decay yields the most precise value and is given by ($m_{\ell}\to0$)
\begin{equation}
\frac{d\Gamma(B\to\pi\ell\nu_{\ell})}{dq^{2}}=\frac{G_{F}^{2}|V_{ub}|^{2}}{192\pi^{3}m_{B}^{3}}\lambda^{3/2}|F_{+}(q^{2})|^{2}\,,
\label{decayrate}
\end{equation}
with  $\lambda=(m_{B}^{2}+m_{\pi}^{2}-q^{2})^{2}-4m_{B}^{2}m_{\pi}^{2}$ and $q^{2}$ the invariant mass of the dilepton pair. $F_{+}(q^{2})$ is the vector FF encoding the $B\to\pi$ transition, the main source of uncertainty in the $|V_{ub}|$ extraction. 

The leptonic differential branching ratio distribution have been measured by {\textit{BABAR}}~\cite{delAmoSanchez:2010af} and BELLE~\cite{Ha:2010rf}. 
This allows us~\cite{VubSergiPere} to extract the $|V_{ub}|$ from a simultaneous fit to the measured $q^{2}$ spectra and lattice simulations on the FF shape at large $q^{2}$ obtained by the HPQCD Coll. in 2006 \cite{Dalgic:2006dt} and by the MILC Coll. in 2008 and 2015~\cite{Bailey:2008wp}.

\begin{wrapfigure}{l}{0.59\textwidth}
\centering
\includegraphics[width=0.58\textwidth]{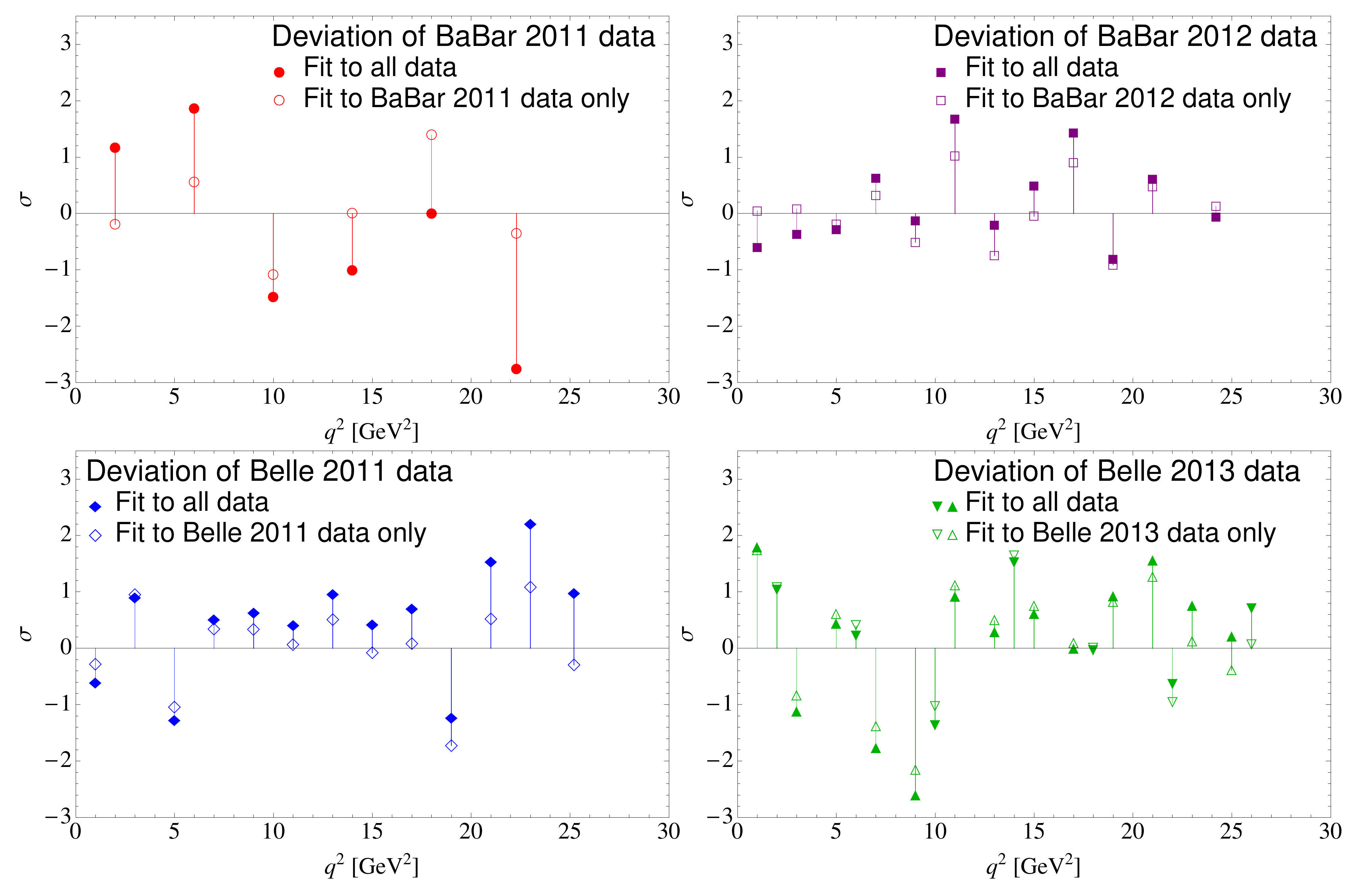}\\
\includegraphics[width=0.58\textwidth]{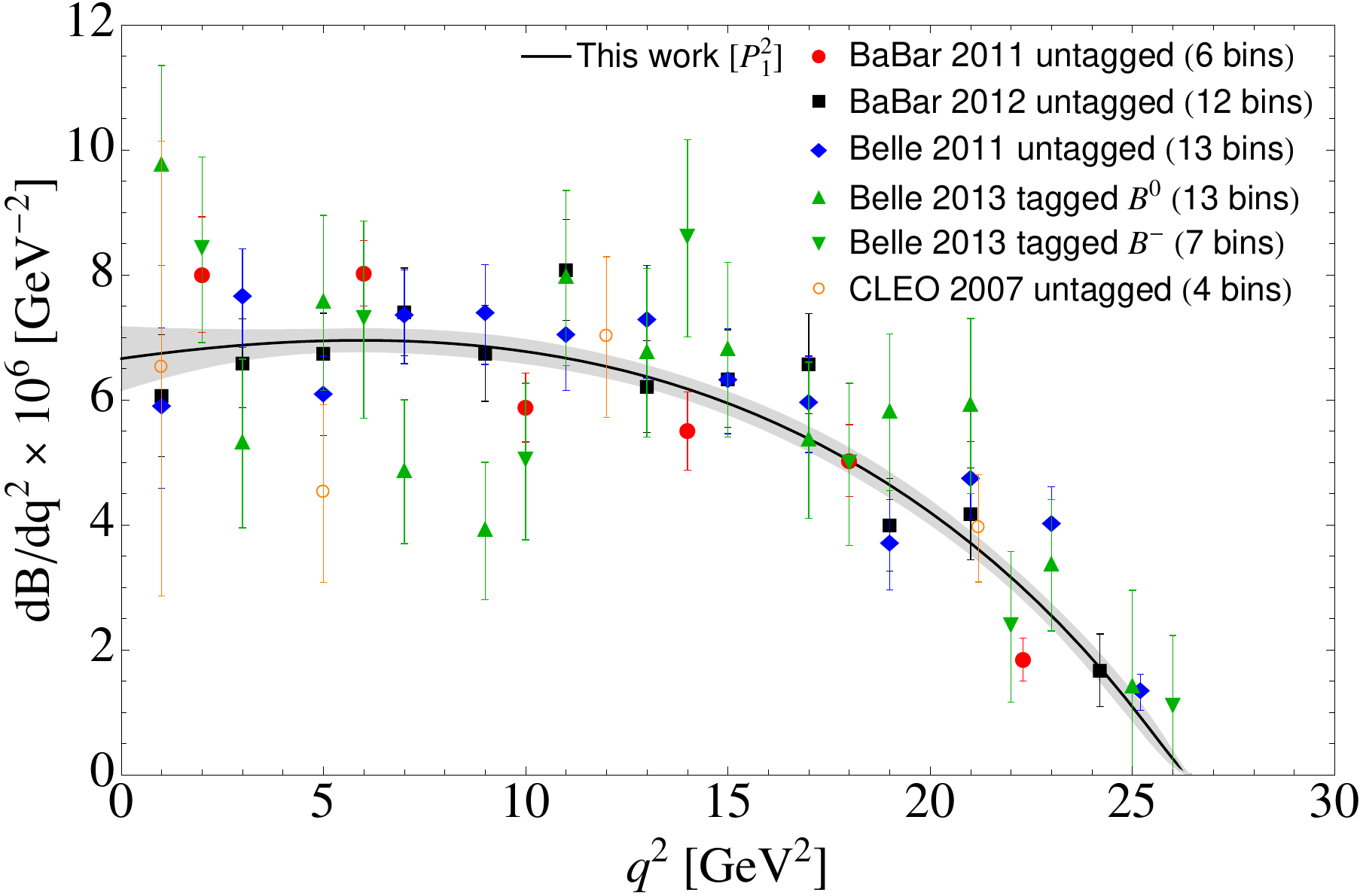}
\caption{{\bf Left:} Deviation, in $\sigma$, of each experimental datum with respect to our combined and individual fits. {\bf Right:} Simultaneous fit to BABAR, BELLE and CLEO data on the $B\to\pi\ell\nu_{\ell}$ from a $\chi^{2}$ minimization with a $P_{1}^{2}(q^{2})$ (black solid line).}
\vspace{-1cm}
\label{Plot_Fit_spectra}
\end{wrapfigure}

For the $B\to\pi\ell\nu_{\ell}$ decay, the lowest threshold appears at $q^{2}=s_{th}=(m_{B}+m_{\pi})^{2}$, above the available kinematical energy range, $0<q^{2}<(m_{B}-m_{\pi})^{2}$, of the decay. 
This explains why, as a first approximation, it has been a common use to consider a VMD model. VMD and di-polar extensions  are elements of the general PA sequence, Eq.~(\ref{PadeApproximant}). 
Since $F_{+}(q^{2})$ is a Stieltjes function higher-order terms in the PA sequence are important to get insights on the analytical structure of the FF and to explore unitary of experimental data. 

We fit with $P_{1,2,3}^{N}(q^2)$ reaching $N=3$ (see example of $P^2_1$ in Fig.~\ref{Plot_Fit_spectra}, lower panel). 
Beyond that, a detailed scrutiny of pole positions in combination with residues of the $\chi^2$ allows us to determine whether a datum satisfies or not unitary constrain (cf. Fig.~\ref{Plot_Fit_spectra}, upper panel). This improves both the quality of the fit and the determination of $|V_{ub}|$~\cite{VubSergiPere}.

\vspace{-0.5cm}
\section{Conclusions and Outlook}

Hadronic form factors are a good laboratory to study the properties of mesons. Their interest goes, however, much beyond the mesons themselves as they play a key role on precision calculations of Standard Model observables at low energies where hadronic contributions are the cornerstone of the error evaluation. We propose the method of Pad\'e approximants as a \underline{toolkit} to analyze them. The method is easy, systematic, user friendly, and can be improved upon by including new data. Provides, as well, information about the underlying structure of the FF and can be used to extrapolate experimental information to extract the low-energy parameters of the FF together with their asymptotic limits. 

The most relevant feature of the method here described is their excellent performance as an interpolation tool thanks to its ability to impose unitary requirements. As such, it is a most compelling method to provide an accurate parameterization for the FF in the whole SL region. Since the approximants can as well penetrate into the TL region below the first resonance, precise experimental data can be easily incorporated. 
In this regard, our method provides an accurate data-driven and model-independent result consistent with the well-known QCD features at high and low energies.

\end{document}